# Crystal and Magnetic Structure of Polar Oxide HoCrWO$_6$

C. Dhital[1*], D. Pham[2], T. Lawal[2], C. Bucholz[1], A. Poyraz[3], Q. Zhang[4], R. Nepal[5], R. Jin[5], R. Rai[6]

[1]Department of Physics, Kennesaw State University, Marietta, GA, 30060, USA

[2] College of engineering, Kennesaw State University, Marietta, GA, 30060, USA

[3]Department of Chemistry and Biochemistry, Kennesaw State University, Kennesaw, GA, 30144, USA

[4]Neutron Scattering Division, Oak Ridge National Laboratory, Oak Ridge, TN, 37830, USA

[5]Department of Physics and Astronomy, Louisiana State University, Baton Rouge, LA, 70803, USA

[6]Department of Physics, SUNY Buffalo State, Buffalo, New York 14222, USA

## Abstract

Polar magnetic oxide HoCrWO$_6$ is synthesized and its crystal structure, magnetic structure, and thermodynamic properties are investigated. HoCrWO$_6$ forms the polar crystal structure (space group *Pna*2$_1$ (#33)) due to the cation ordering of W$^{6+}$ and Cr$^{3+}$. There is an antiferromagnetic transition at $T_N$ = 24.5 K along with the magnetic entropy change (~5 J.Kg.$^{-1}$K$^{-1}$ at 70 kOe). Neutron diffraction measurement indicates that both Cr and Ho sublattices are ordered with the moment of 2.32(5)$\mu_B$ and 8.7(4)$\mu_B$ at 2 K, respectively. While Cr forms A-type collinear antiferromagnetic (AFM) structure with magnetic moment along the *b* axis, Ho sublattice orders in a non-coplanar AFM arrangement. A comparison with isostructural DyFeWO$_6$ and DyCrWO$_6$ indicates that the magnetic structure of this family of compounds is controlled by the presence or absence of $e_g$ electrons in the transition metal sublattice.

*cdhital@kennesaw.edu

1. ## Introduction

Polar (noncentrosymmetric and achiral) magnetic oxides are systems of interest due to their fascinating fundamental properties desirable for applications such as ferroelectricity, piezoelectricity, magnetoelectricity, or multiferroicity (1-16). These physical properties are especially attractive for advanced electronic devices

(4, 12-14). In the case of polar magnetic compounds, the emphasis is given to the investigation of possible multiferroicity (1-4, 13-14), due to the possibility of tuning magnetic properties by electric field and vice versa (1-5,17,18). However, the multiferroic materials studied so far have either low magnetic ordering temperature or weak magnetoelectric coupling (1-5,17-23). Therefore, there is increasing need for the discovery of new polar magnetic materials. The magnetoelectric coupling is known to be stronger in type II multiferroic materials compared to type I materials (1,3,5,17-23). In type II materials, the non-collinear magnetic ordering (e.g. cycloidal) breaks the space inversion symmetry of the lattice through inverse Dzyalloshinkii-Moriya (DM) interaction or exchange striction (1-5), and introduces ferroelectric ordering thereby increasing the magnetoelectric coupling (1,3,5,17-23). Therefore, it is essential to understand the magnetic structure and magnetic behavior of polar magnetic materials in order to tune magnetoelectric coupling strength.

Recently, type II magnetoelectric multiferroic properties have been reported in some members of the $RFeWO_6$ (R=Y or Dy, Tb, Eu) family that are polar even in the paramagnetic state (24). Therefore, this material family might provide new insight into how the magnetoelectric coupling is affected by the presence of non-centrosymmetricity in the paramagnetic state. Previous studies (24-26) were performed on $YCrWO_6$, $YFeWO_6$, $DyCrWO_6$, $DyFeWO_6$, $EuFeWO_6$, $TbFeWO_6$. Even though both $RCrWO_6$ and $RFeWO_6$ (R = Dy, Y) compounds have the same polar crystal structure, ferroelectric polarization was only observed in the magnetic state of $RFeWO_6$ (24-26) through magnetoelectric coupling. The absence of electric polarization in Cr-based compounds was explained based on the quasi collinear arrangement of Cr moments (26) and possible absence of magnetoelectric coupling. However, such explanation was based only on the neutron diffraction study on a single compound $DyCrWO_6$ (26). We have extended this line of inquiry to isostructural compound $HoCrWO_6$ to investigate with focus on (I) the stability of polar crystal structure with the change in size of the rare earth ion, and (II) the role of $e_g$ electrons and resulting exchange interactions for the spin arrangement in the transition metal sublattice.

Here, we present the results from neutron diffraction, magnetization and specific heat measurements on $HoCrWO_6$. We discuss our results with reference to previous work on $DyCrWO_6$, $DyFeWO_6$ and $YCrWO_6$. The results from our work and comparison with previous work indicate that the polar crystal structure is stable in $HoCrWO_6$ and the absence of $e_g$ electrons in Cr sublattice results in the magnetic structure of $HoCrWO_6$ similar to that of $DyCrWO_6$ but different to that of $DyFeWO_6$.

## 2. Experimental Section

Polycrystalline $HoCrWO_6$ was synthesized by the solid-state reaction method. Stoichiometric amounts of $Ho_2O_3$ (preheated at 900 °C for 12 h in air), $Cr_2O_3$, and $WO_3$ were mixed, ground, and heated at 1180 °C for 18 h in air with intermittent grindings. Preliminary phase purity was determined using X-ray diffraction (XRD). XRD patterns were collected on a Rigaku Miniflex XRD instrument (Cu Kα radiation λ= 1.5406 Å, 40 kV, and 20 mA) equipped with D/teX Ultra detector. Magnetization measurements were performed using a SQUID magnetometer (SCM5) in National High Magnetic Field Laboratory Florida, and Dynacool vibrating sample magnetometer (VSM) system at SUNY Buffalo State. The specific heat was measured using a Physical Property Measurement System (Quantum Design). Neutron diffraction measurements were performed using the time of flight (TOF) technique in POWGEN instrument in spallation neutron source (SNS) at Oak Ridge National Laboratory using the neutron spectrum of central wavelength 1.5 Å. The crystal and magnetic structure were refined using Fullprof (27) and GSAS-II software (35).

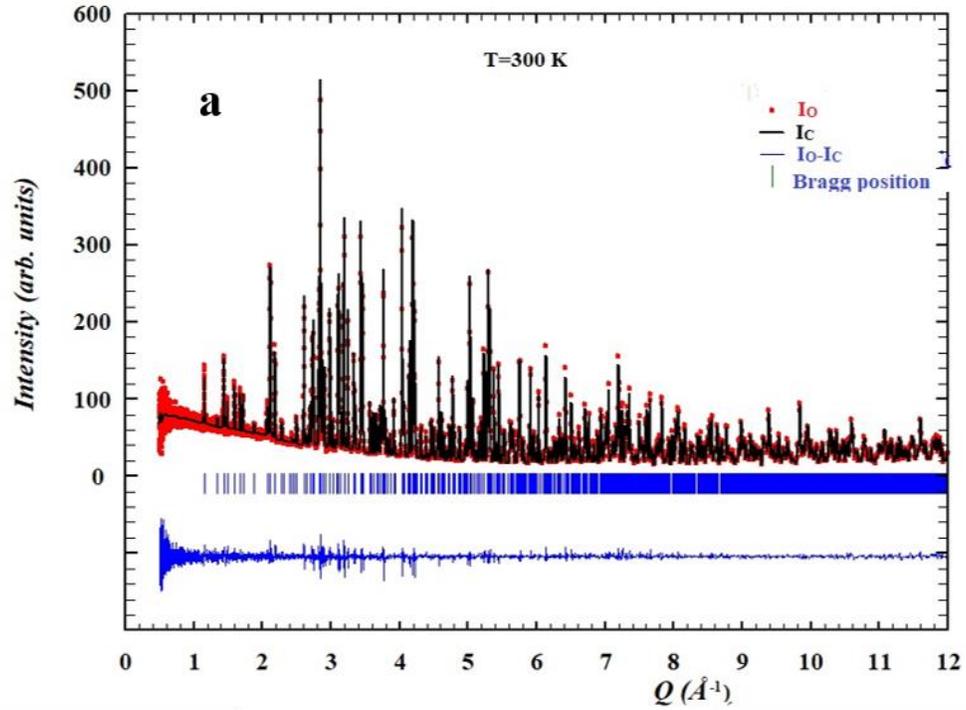

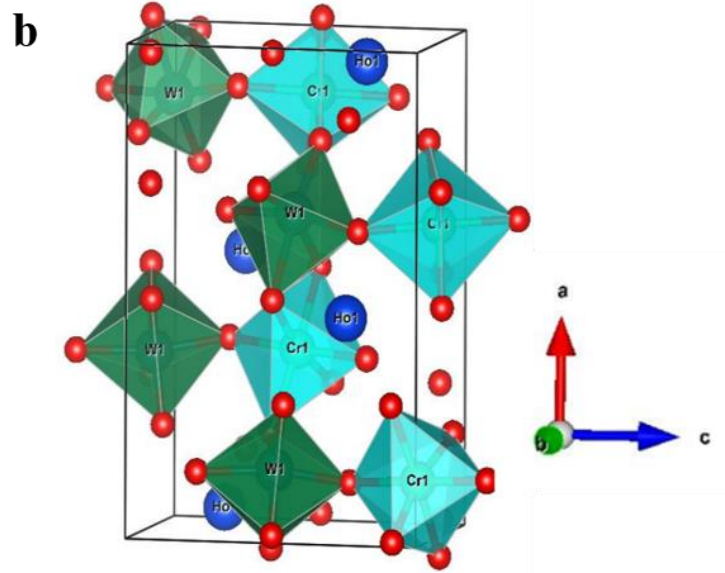

Figure 1 Structural analysis of HoCrWO$_6$. (a) Refinement of neutron powder diffraction data at 300 K. The red points are observed intensity ($I_o$), black curves are calculated intensity ($I_C$), the blue vertical lines are Bragg positions and the blue horizontal line represents the difference between observed and calculated intensities ($I_o$-$I_C$). (b) Crystal structure of HoCrWO$_6$ viewed along the $b$ axis. The atoms: Ho (blue), Cr (cyan), W(green) and O (red) are denoted by spheres.

## 3. Results and Discussion
### 3.a. Crystal structure determination

The room temperature ($T=300$ K) TOF neutron diffraction data was corrected for neutron absorption using Lorentz correction and refined using the Fullprof software. The profile parameters were fixed to the instrument resolution and isotropic thermal factors, the atomic coordinates and the lattice parameters were refined. The crystal structure was determined to be in the polar orthorhombic space group $Pna2_1$ (#33) with the lattice parameters $a = 10.8724(2)$ Å, $b = 5.1661(1)$ Å, $c = 7.3096(1)$ Å, $V = 410.57(1)$ Å$^3$ and $Z = 4$. The observed intensity ($I_o$), calculated intensity ($I_C$), the difference ($I_o$-$I_C$), and the Bragg positions are shown in Figure 1. The good quality of refinement is indicated by the $R$-factor of 3.63 and RF-factor of 2.78. HoCrWO$_6$ has three-dimensional crystal structure consisting of edge sharing pair of WO$_6$ and CrO$_6$ octahedra similar to DyCrWO$_6$ and YCrWO$_6$, as shown in Figure 1b. This result is consistent with the previously reported polar crystal structure of YCrWO$_6$ (25) because Y$^{3+}$ and Ho$^{3+}$ have almost the same values of ionic radii. The structure of HoCrWO$_6$ can be taken to be derived from the euxenite ($Pbcn$) type Ho(TaTi)O$_6$ or aeschynite ($Pnma$) type Dy(TaTi)O$_6$ (28). In the euxenite or aeschynite type oxides, the disordered Ta or Ti ions occupy unique crystallographic site (*8d*) randomly (28). In the case of HoCrWO$_6$, the *8d* crystallographic site splits into two *4a* octahedral sites, where W$^{6+}$ and Cr$^{3+}$ are ordered. The origin of such cation ordering lies in the difference in the octahedral distortion parameter $\Delta = \frac{1}{6}\sum_i \left(\frac{d_i - d_m}{d_m}\right)^2$ between CrO$_6$ and WO$_6$ octahedra. Here, $d_i$ is the individual bond length and $d_m$ is the average bond length between Cr$^{3+}$ or W$^{6+}$ cation and oxygen anion ($\Delta = 0.0025$ for WO$_6$ and $\Delta = 0.00013$ for CrO$_6$). The W-O bond distances vary from 1.82 to 2.08 Å, Cr-O bond distances vary from 1.95 to 2.006 Å, and Ho-O distances vary from 2.29 Å to 2.46 Å. The average valence of each element was also calculated using the bond valence sum (BVS) calculations. The BVS calculation is consistent with the formal oxidation state of Cr and Ho. The structural parameters and BVS at 300 K, bond lengths, and bond angles are presented in Tables I-III, respectively, in supplementary materials (29).

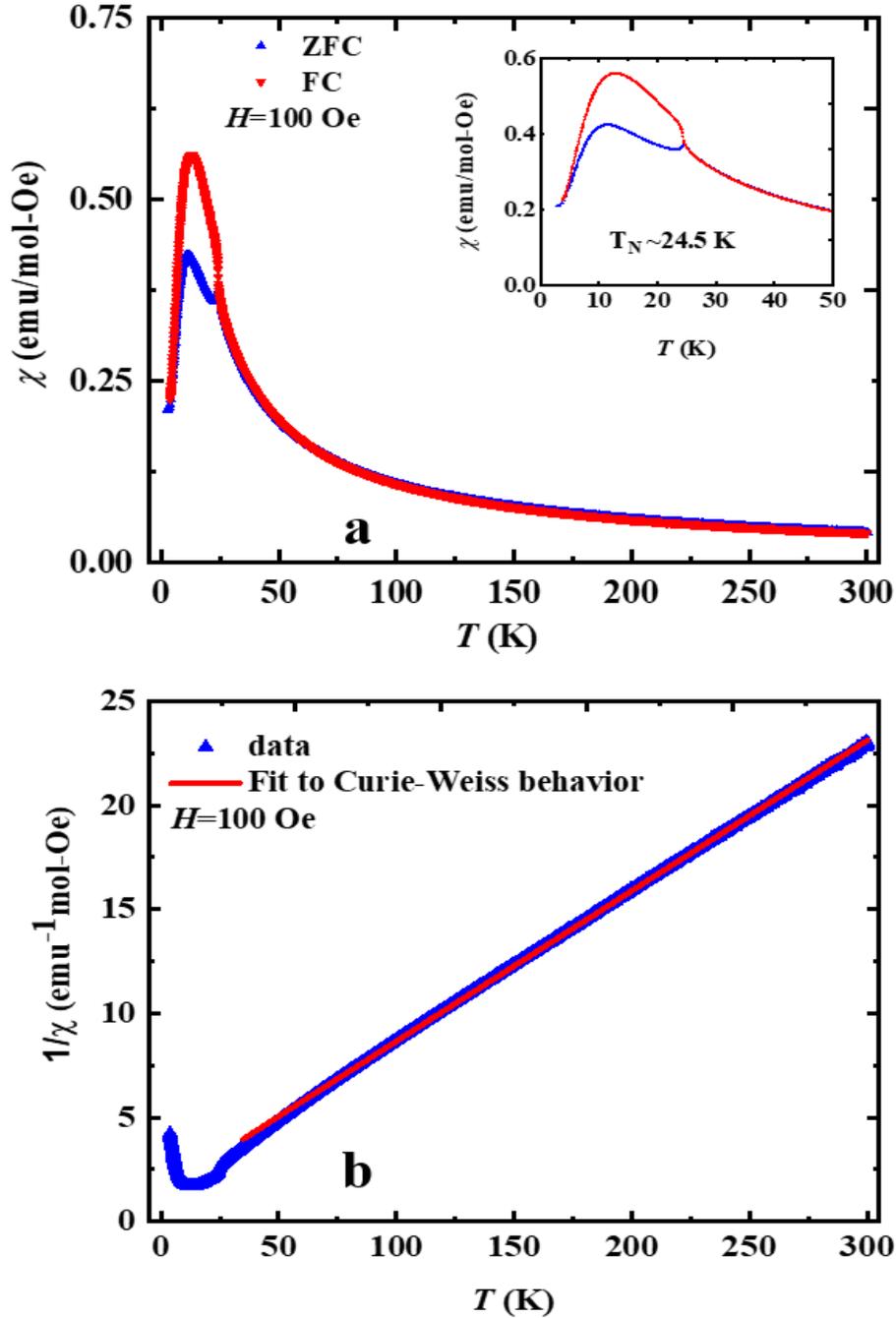

Figure 2 Magnetic properties of HoCrWO$_6$. (a) The ZFC and FC magnetic susceptibility, $\chi$, versus temperature, $T$, measured at $H=100$ Oe. Inset: zoomed in $\chi$ versus $T$ curve around the magnetic transition. (b) The inverse susceptibility is plotted against temperature and fitted to $\frac{1}{\chi} = \left(\frac{1}{C}\right)T - \left(\frac{\theta}{C}\right)$, where $C$ is the Curie-Weiss constant and $\theta$ is the Curie-Weiss temperature. The red line is the fit to data in the temperature range 300 K to 35 K.

### 3.b. Magnetic properties measurements

Figure 2a shows the zero-field cooled (ZFC) and field cooled (FC) magnetic susceptibility ($\chi$), under the field of $H = 100$ Oe, as a function of temperature, $T$. There is a clear anomaly at $T_N = 24.5$ K, as shown in the inset, suggesting long range ordering. Upon further cooling $\chi(T)$ further increases and exhibits a broad maximum around 13 K. The ZFC and FC curves meet at low temperature most probably due to spin reorientation involving Ho spins and due to increased antiferromagnetic coupling between Ho and Cr spins at low temperature. The $\chi$-$T$ curve for HoCrWO$_6$ is similar to that of DyCrWO$_6$ (24), in which both the rare earth Dy and Cr sublattices were found to be ordered below $T_N$. Fitting of magnetic susceptibility $\chi$ to the Curie-Weiss law (29), $\frac{1}{C} = \left(\frac{1}{C}\right)T - \left(\frac{\theta}{C}\right)$ (Fig. 2b) in the temperature range of 300 K to 35 K results in the Curie-Weiss constant $C = 13.85$ emu-K/mol-Oe and the Curie-Weiss temperature $\theta = -19.8$ K. The negative Curie-Weiss temperature indicates antiferromagnetic ordering at $T_N$. The small deviation of $\theta$ from $T_N$ indicates negligible magnetic frustration. From the Curie-Weiss constant, the effective moment is calculated to be 10.52$\mu_B$/f.u. (f.u represents the formula unit). This is smaller than the expected sum of free ion moments of Ho$^{3+}$ (~10.4$\mu_B$) and Cr$^{3+}$ (~3.4$\mu_B$) (30).

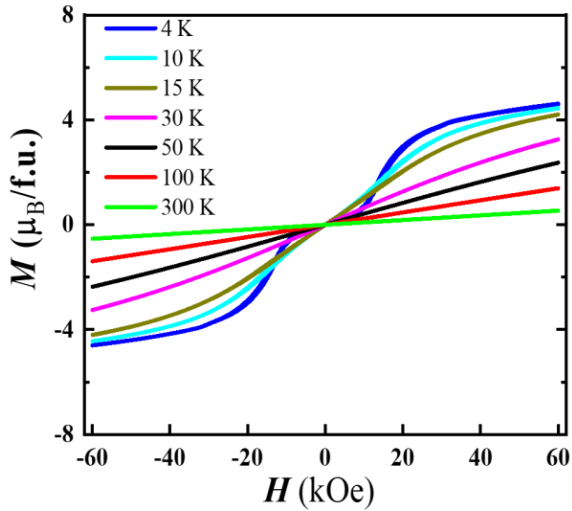

Figure 3 Magnetization, *M,* as function of magnetic field, *H*, for HoCrWO$_6$.

The magnetic field dependence of the magnetization *M(H)* at different temperatures are shown in Figure 3. At $T > T_N$, the magnetization exhibits linear *H* dependence, consistent with the paramagnetic phase. Non-linear *M* versus *H* is observed below $T_N$, suggesting canted antiferromagnetic phase. At 4 K (blue curve), there is a step like behavior in *M(H)* curve, indicating the metamagnetic transition of probably $Ho^{3+}$ moments. The value of the magnetization at 60 kOe and 4 K (~$4\mu_B$/f.u) is less than that predicted by the Curie-Weiss law (30). While it shows nearly saturated ferromagnetic type behavior at 4 K in higher field, however, the saturation magnetization is significantly smaller than the expected sum of Cr and Ho free ion moments. This is unusual in such an insulating system. The primary reason for such observation is the powder averaging of easy and hard axis magnetization due to polycrystalline nature of sample. Another less likely possibility is the partial quenching of orbital component due to crystal electric field and associated magneto crystalline anisotropy (31).

Since the magnetic state changes with magnetic field at low temperatures, we have also investigated the magnetocaloric property of $HoCrWO_6$ by calculating the change in magnetic entropy. By measuring magnetization versus temperature at different magnetic fields and rearranging the data, we have estimated the change in the magnetic entropy using relation (32)

$$\Delta S(T, 0 \rightarrow H) = \int_0^H \left(\frac{\partial M}{\partial T}\right)_H dH$$

For the magnetization data taken at discrete magnetic fields and temperatures, the above equation can be integrated numerically as (32)

$$\Delta S = \sum_i \frac{M_{i+1} - M_i}{T_{i+1} - T_i} \Delta H_i$$

where $M_{i+1}$ and $M_i$ are the measured magnetizations at temperatures $T_{i+1}$ and $T_i$, respectively. The magnetization and its temperature derivative as a function of temperature are shown in Figures 4a and 4b, respectively. The change in magnetic entropy $\Delta S_M$, with respect temperature and field, is presented in Figure 5a. It is clear from Figure 5a that $HoCrWO_6$ shows double peaks in -$\Delta S_M$, a relatively sharp peak around 24 K, and the second, field dependent broad peak at lower temperatures. The two peaks start to merge together with increasing field and form a single broad peak at 70 kOe. A change in magnetic entropy (~5 J.kg.$^{-1}$K.$^{-1}$) occurs around the magnetic transition at a field of 70 kOe. The presence of double peaks increases the working temperature range for magnetic refrigeration and the refrigerant capacity (RC). Quantitatively, the RC value is calculated as RC = $(\Delta S_M)(\Delta T_{cycl})$ (32, 33) where $\Delta S_M$ is the value of entropy change at full width half maximum (FWHM) and $\Delta T_{cycl} = T_{hot}-T_{cold}$ is the temperature difference between two end points at FWHM. Taking $\Delta S_M = 2.5$ J.kg.$^{-1}$K.$^{-1}$, and $T_{cycl} = T_{hot}-T_{cold} = 50$ K -10 K = 40 K, RC is 100 J.kg.$^{-1}$. Such a value of RC is comparable to other members of this family and manganites (32). There is also a sign change of $\Delta S_M$ at low temperatures (below 8 K) and at field of 10 kOe. The field of 10 kOe corresponds to the onset of metamagnetic transition (see Fig 3). Therefore, the sign change of entropy at 10 kOe is related to the metamagnetic transition. We have also measured the specific heat at $H = 0$, 10, and 70 kOe as presented in Figure 5b. The zero-field specific heat (see the inset of Figure 5b) shows the λ-type anomaly at $T_N$ and a broad feature centered around 10 K. This broad feature is in the temperature range where there is a downturn in the magnetic susceptibility (Fig. 2a). We have also compared the magnetic entropy change as calculated from the magnetization and specific heat (See supplementary materials Fig. S1). The sign change of the magnetic entropy changes in the low temperature region around 10 kOe field is evident from both methods of calculations (See supplementary Figure S1). This indicates that the sign reversal of entropy change is related to the increase in spin disorder in the vicinity of metamagnetic transition.

### 3.c. Magnetic structure determination

In order to understand the magnetic behavior, we also performed low temperature ($T = 2$ K and $T = 18$ K) neutron diffraction study on HoCrWO$_6$. The low temperature neutron diffraction data along with the refinement are presented in Fig 6. The low temperature ($T < T_N$) diffraction patterns contain extra peaks that were absent at room temperature ($T = 300$ K, Figure 1a). These extra peaks emerge at the low momentum transfer, $Q$, region below the magnetic ordering temperature $T_N$, therefore, are magnetic in origin. Fig. 7a presents the evolution of two low $Q$ magnetic peaks above and below the magnetic transition. Indexing of the extra peaks using Fullprof program gives the commensurate magnetic propagation vector $k = (0, 0, 0)$. The magnetic structure is determined by $k$-subgroupsmag (34) and Rietveld refinement using GSAS-II (35). There are 11 subgroups consistent with the $Pna2_1$ space group and the magnetic propagation vector. Out of these 11 magnetic space groups, the space group $Pna2_1$ (#33.144) best fits our data. The fitting of experimental data performed using GSAS-II is presented in Figs. 6a and 6b. The resulting magnetic structure at 2 K is shown in Fig 7b. The magnetic structure at 18 K is not much different from the 2 K structure except reduced value of moments and slight change in orientation of Ho moments. The resulting magnetic moment values on Cr and Ho sites are presented in Table I.

**Table 1: Components of magnetic moments at 2 K and 18 K**

| T (K) | Atom | $m_x$ ($\mu_B$) | $m_y$ ($\mu_B$) | $m_z$ ($\mu_B$) | $m$ ($\mu_B$) | constrain |
|---|---|---|---|---|---|---|
| 2 | Cr | 0 | -2.32(5) | 0 | 2.32(5) | $m_z=0$, $m_x=0$ |
|   | Ho | -0.91(5) | 5.65(3) | 6.53(2) | 8.7(4) | |
| 18 | Cr | 0 | -2.21(4) | 0 | 2.21(3) | $m_z=0$, $m_x=0$ |
|   | Ho | 0 | 2.27(3) | 2.60(2) | 3.4(3) | $m_x=0$ |

It should be noted that some of the components were constrained when the free refinement gave unreasonable values with significant errors without improving the quality of refinement. The magnetic structure at 2 K is slightly different from that at 18 K as reflected in the *x* component of Ho moment. This evolution of the *x* component of Ho moment should be the reason for the downturn in the magnetization around 12 K and associated magnetic entropy change.

The magnetic structure of HoCrWO$_6$ is similar to that of DyCrWO$_6$ with some minor differences. Both HoCrWO$_6$ and DyCrWO$_6$ have the same magnetic wavevector $k = (0\ 0\ 0)$. In HoCrWO$_6$, the magnetic space group *Pna*2$_1$ (#33.144) best fits our data whereas in DyCrWO$_6$ the magnetic space group *P*112$_1$ was used. In the *Pna*2$_1$ space group there is one Cr and one Ho site whereas in *P*112$_1$ both Cr and Ho sites split into two sites. However, in both HoCrWO$_6$ and DyCrWO$_6$, the arrangement of magnetic moments is in the similar manner.

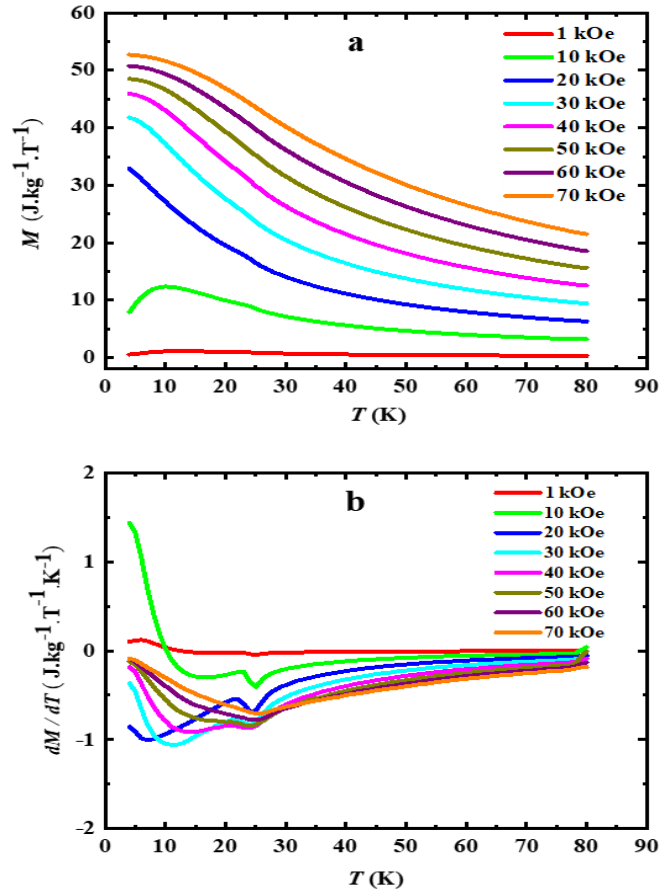

Figure 4. (a) Magnetization vs temperature at indicated fields. (b) Data replotted as d$M$/d$T$ versus $T$.

In HoCrWO$_6$, the closest Cr$^{3+}$ - Cr$^{3+}$ distance is 4.67 Å, the closest Cr$^{3+}$ - Ho$^{3+}$ distance is 3.36 Å and the closest Ho$^{3+}$ - Ho$^{3+}$ distance is 3.79 Å. Given the crystal structure, there cannot be direct exchange interaction between any of these magnetic ions. Since the magnetic structure of HoCrWO$_6$ is similar to that of DyCrWO$_6$, the scenario of strongest Cr$^{3+}$-O-O-Cr$^{3+}$ antiferromagnetic super-super exchange interaction and weakest Ho$^{3+}$-O-Ho$^{3+}$ interaction might also be valid for HoCrWO$_6$ as presented for DyCrWO$_6$ through exchange integral calculation (26,36-42).

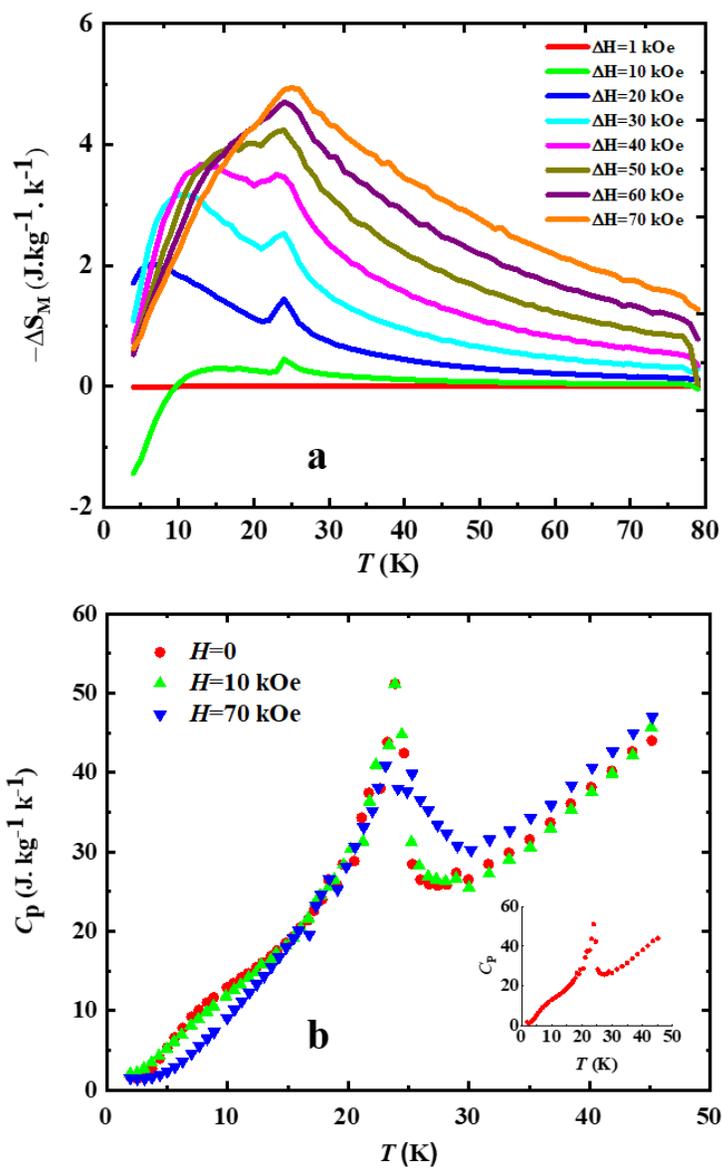

Figure 5. (a) Change in the magnetic entropy as function of temperature for different fields. (b) Heat capacity, $C_p$, vs $T$. Inset shows specific heat measured at zero field.

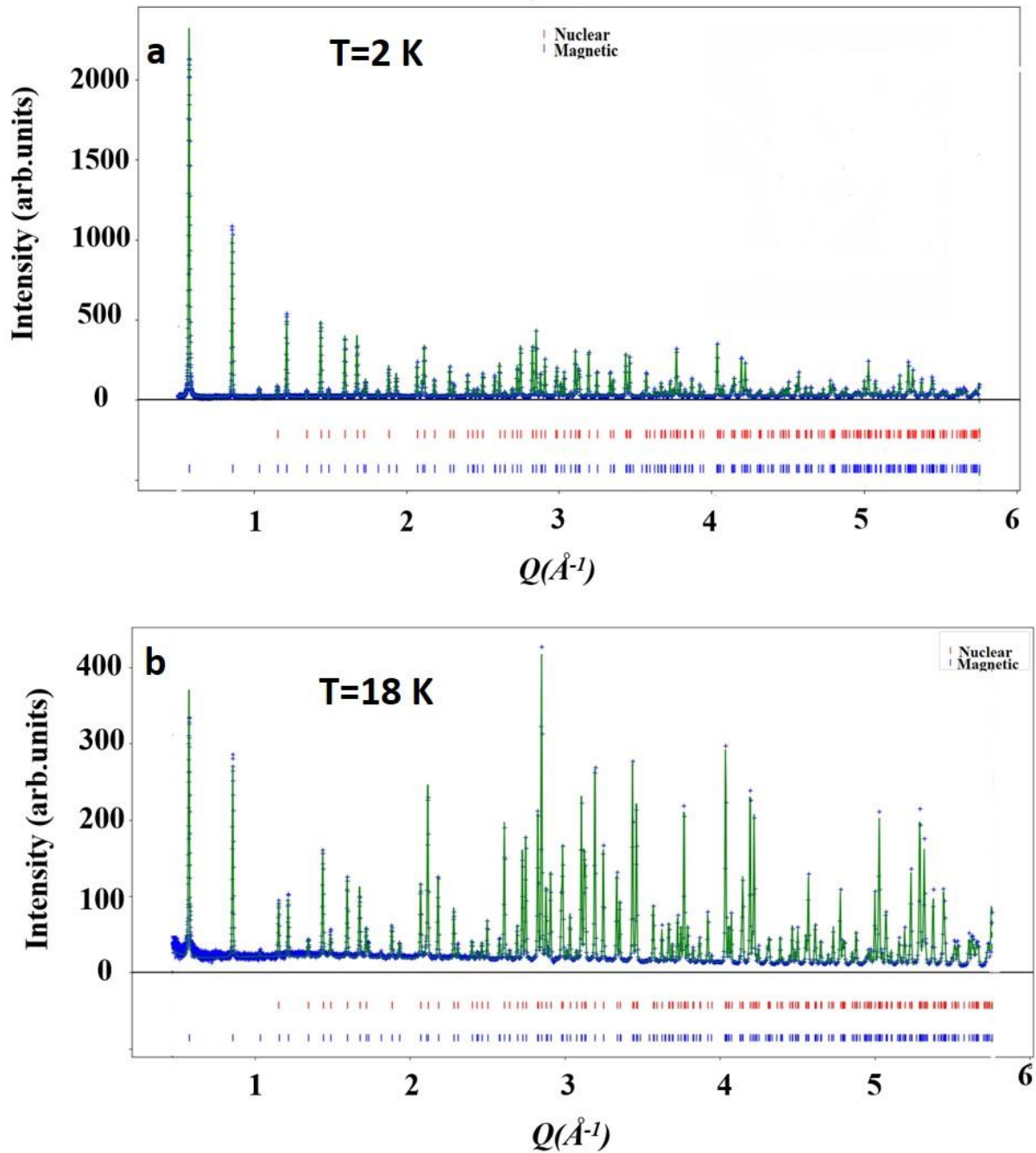

Figure 6. Neutron diffraction data with Rietveld refinement on HoCrWO$_6$. (a) at 2 K and (b) at 18 K. The blue points represent actual data and green line represent the Rietveld fitting to the data. The red marks are allowed nuclear reflections and blue marks are allowed magnetic reflections.

In this work, we have performed neutron diffraction measurements at 2 K and 18 K where both Ho and Cr sublattices are ordered, similarly to previous work on DyCrWO$_6$ and DyFeWO$_6$ (24,26). In previous work (24, 26), it was observed that both rare earth and transition metal sublattices order below $T_N$ and the magnetic moment on the rare earth site increases progressively as temperature is lowered. It should be mentioned that powder diffraction gives the average intensity and becomes insensitive to slight change in orientation of the moments. Therefore, future single crystal neutron diffraction would be ideal to find the details of magnetic structure.

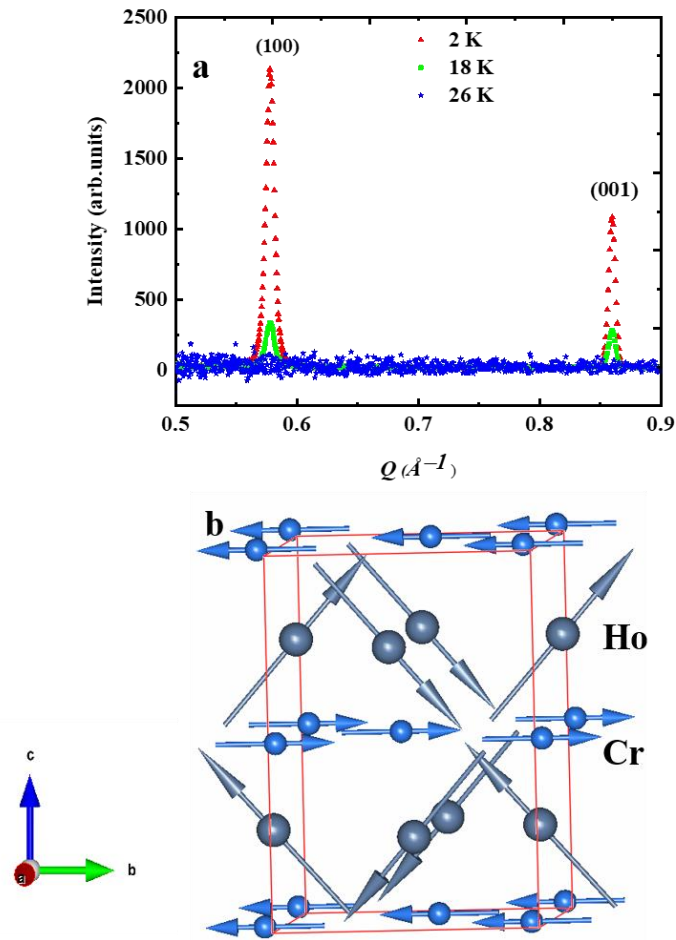

Figure 7. Magnetic structure determination of HoCrWO$_6$ .(a) Comparison of magnetic peaks of HoCrWO$_6$ at different temperatures (b) Magnetic structure of HoCrWO$_6$ at T=2 K viewed along *a*-axis. Red lines represent boundary of unit cell.

Although the magnetic structure of HoCrWO$_6$ and DyCrWO$_6$ are similar, the magnetic structure of DyFeWO$_6$ is different (24). In DyFeWO$_6$, both Fe and Ho spins are in non-collinear arrangement whereas, in DyCrWO$_6$ and HoCrWO$_6$, the Cr spins are arranged in almost collinear ($m_x \ll m_y$, $m_z \approx 0$) spin arrangement with the A-type configuration. This difference in the spin arrangement between Fe and Cr compounds indicates the role of $e_g$ electrons in Fe$^{3+}$ that are absent in Cr$^{3+}$ in such octahedral crystal environment. The presence or absence of $e_g$ electrons can cause the change in the strength of Cr$^{3+}$-O-Ho$^{3+}$ superexchange interaction and hence the magnetic structure.

## 4. Conclusion

The cation ordering of Cr$^{3+}$ and W$^{6+}$ in connected CrO$_6$ and WO$_6$ octahedra breaks the space inversion symmetry of the lattice resulting in a polar crystal structure of HoCrWO$_6$ similar to other R(Fe,Cr)WO$_6$ compounds. The Curie-Weiss behavior indicates an antiferromagnetic ordering with very little frustration indicated by the frustration parameter ($\theta/T_N$) being close to unity. HoCrWO$_6$ exhibits double magnetocaloric peaks at low temperatures with the refrigerant capacity $RC = 100$ J, kg$^{-1}$ corresponding to the antiferromagnetic transition at $T_N$ and Ho spin reorientation below 12 K. At 2 K, both Cr and Ho sublattices are ordered with the ordered moments of 2.32(5)$\mu_B$ and 8.7(4)$\mu_B$, respectively. The values of ordered moments decrease with increasing temperature. At 18 K, Cr and Ho are 2.21(3)$\mu_B$ and 3.4(3)$\mu_B$, respectively. The Cr spins are arranged in collinear ferromagnetic arrangement along the *b* axis and the ferromagnetic planes are stacked antiferromagnetically along the *c* axis, i.e., the A-type antiferromagnetic ordering. The Ho spins are arranged in a non-collinear spin arrangement. Similar to DyCrWO$_6$, the magnetic interaction strength can most probably be arranged as Cr-O--O-Cr > Cr-O-Ho > Ho-O-Ho. A comparison between HoCrWO$_6$, DyCrWO$_6$ and DyFeWO$_6$ is presented in Table II. Comparing these isostructural compounds, the spin arrangement of the transition metal sublattice is determined by the presence or absence of $e_g$ electrons.

The space group *Pna*2$_1$ with polar point group *mm*2 allows spontaneous polarization along *c*-axis even at room temperature. However, no polarization was observed in non-magnetic state in other members of this family due to polycrystalline sample and the requirement of large polling electric field. In the magnetic state, the observation of polarization depends upon the strength of magnetoelectric coupling. The magnetoelectric coupling in turn depends upon the spin arrangement in the transition metal sublattice. Future polarization studies in magnetic fields are desirable in this compound to understand the consequence of collinear spin arrangement in the strength of magnetoelectric coupling.

In summary, with the synthesis and characterization new polar compound HoCrWO6, our work allows the comparison of crystal and magnetic structure of different members of RMWO$_6$ family. Our work also provides future avenue to tune the magnetoelectric coupling strength and the transition temperatures by suitable choice of rare earth and the transition metal ions in RMWO$_6$ family.

**Table II:** Comparison of the parameters within the RMWO$_6$ (R=Ho, Dy and M=Fe, Cr) family: *a,b,c* are lattice parameters, $T_N$ is the antiferromagnetic ordering temperature, *Ms* is the value of magnetization at low temperature (2 K-4 K) at 60 kOe field.

| Compound | *a,b,c* (Å) T=300 K | $T_N$ (K) | Nature of ordering in *M* site | Nature of ordering in *R* site | *Ms* ($\mu_B$) (magnetization) | Reference |
|---|---|---|---|---|---|---|
| HoCrWO$_6$ | *a*=10.8724(2) *b* = 5.1661(1) *c* = 7.3096 (2) | 24.5 | Quasi-collinear $m_y>>m_x$, $m_z=0$ | Non collinear $m_x=0$, $m_y \sim m_z$ | ~4.6 | This work |
| DyCrWO$_6$ | *a*=10.87855(2) *b* = 5.18328(1) *c* = 7.32152(2) | 25 | Quasi-collinear $m_y>>m_x$, $mz \approx 0$ | Non-collinear $m_x<m_y \approx m_z$ | ~5 | Ref.[26] |
| DyFeWO$_6$ | *a* = 10.97992(2) *b* = 5.18849(1) *c* = 7.34824(1) | 18 | Non-collinear (based on magnetic structure) | Non-collinear (based on magnetic structure) | ~5 | Ref.[24] |

## 5. Supplementary Material

The supplementary material contains the structural information such as bond lengths, bond angles, the comparison of magnetic entropy calculation from heat capacity and magnetization data.

**Acknowledgements**

The work at Kennesaw State University was supported by faculty startup grant. Work at SUNY Buffalo State was supported by the National Science Foundation (Grant No. DMR-1406766). RN and RJ are supported by by the U.S. Department of Energy (DOE) under EPSCoR Grant No. DE-SC0016315. A portion of this research used resources at the Spallation Neutron Source, a DOE Office of Science

User Facility operated by the Oak Ridge National Laboratory. A portion of this work was performed at the National High Magnetic Field Laboratory, which is supported by National Science Foundation Cooperative Agreement No. DMR-1644779 and the State of Florida.

# Crystal and Magnetic Structure of Polar Oxide HoCrWO$_6$

C. Dhital[1*], D. Pham[2], T. Lawal[2], C. Bucholz[1], A. Poyraz[3], Q. Zhang[4], R. Nepal[5], R. Jin[5], R. Rai[6]

[1]Department of Physics, Kennesaw State University, Marietta, GA, 30060, USA

[2] College of engineering, Kennesaw State University, Marietta, GA, 30060, USA

[3]Department of Chemistry and Biochemistry, Kennesaw State University, Kennesaw, GA, 30144, USA

[4]Neutron Scattering Division, Oak Ridge National Laboratory, Oak Ridge, TN, 37830

[5]Department of Physics and Astronomy, Louisiana State University, Baton Rouge, LA, 70803

[6]Department of Physics, SUNY Buffalo State, Buffalo, New York 14222, USA

We present structural parameters, bond lengths and bond angles for HoCrWO$_6$.

**Table S1:** Structure parameters obtained from neutron diffraction data on HoCrWO$_6$ at 300 K Space group: *Pna2$_1$*; $a = 10.8724(2)$ Å, $b = 5.1661(1)$ Å, $c = 7.3096(2)$ Å, $\alpha = \beta = \gamma = 90°$, $V = 410.57(1)$ Å$^3$; $\chi^2 = 1.05$; Bragg R factor = 3.63, R$_f$ factor = 2.78. Wyckoff positions (x, y, z), occupancy (occ), isotropic thermal factor (U$_{iso}$) and site index (site) are presented in the Table below

| Atom | x | y | z | Occ. | U$_{iso}$ | Site |
|---|---|---|---|---|---|---|
| W1  | 0.35440 | 0.44400 | 0.00400 | 1.000 | 0.001 | 4a |
| Ho1 | 0.04310 | 0.45460 | 0.25000 | 1.000 | 0.002 | 4a |
| Cr1 | 0.13250 | 0.95300 | 0.99900 | 1.000 | 0.009 | 4a |
| O1  | 0.97290 | 0.76840 | 0.04300 | 1.000 | 0.003 | 4a |
| O2  | 0.52140 | 0.25600 | 0.96100 | 1.000 | 0.006 | 4a |
| O3  | 0.21420 | 0.61700 | 0.06300 | 1.000 | 0.007 | 4a |
| O4  | 0.29150 | 0.12700 | 0.93500 | 1.000 | 0.005 | 4a |
| O5  | 0.14310 | 0.05950 | 0.25500 | 1.000 | 0.006 | 4a |
| O6  | 0.11910 | 0.82940 | 0.74400 | 1.000 | 0.005 | 4a |

**Table S2**: Bond lengths and bond valence sums (BVS) for select bonds for HoCrWO$_6$ at 300 K

| Bond | Bond length (Å) | Bond valence sum (BVS) |
|---|---|---|
| W1-O1 | 1.987(2) | |
| W1-O2 | 2.083(2) | |
| W1-O3 | 1.818(9) | |
| W1-O4 | 1.845(4) | 6.11 (W$^{6+}$) |
| W1-O5 | 1.915(6) | |
| W1-O6 | 1.873(8) | |
| Ho1-O1 | 2.438(1) | |
| Ho1-O2 | 2.301(1) | |
| Ho1-O3 | 2.456(2) | |
| Ho1-O4 | 2.419(9) | 3.19 (Ho$^{3+}$) |
| Ho1-O5 | 2.312(9) | |
| Ho1-O6 | 2.294(5) | |
| Cr1-O1 | 2.006(1) | |
| Cr1-O2 | 1.948(4) | |
| Cr1-O3 | 2.005(2) | |
| Cr1-O4 | 2.003(8) | 2.99 (Cr$^{3+}$) |
| Cr1-O5 | 1.953(9) | |
| Cr1-O6 | 1.975(7) | |

**Table S3:** Select bond angles for HoCrWO$_6$ at 300 K

| Bond | Bond angle (in degree) |
|---|---|
| Cr1-O1-W1 | 100.5 (7) |
| Cr1-O2-W1 | 99.2 (7) |
| Cr1-O3-W1 | 137.8(11) |
| Cr1-O4-W1 | 130.9(9) |
| Cr1-O5-W1 | 145.4(7) |
| Cr1-O6-W1 | 140.3(7) |

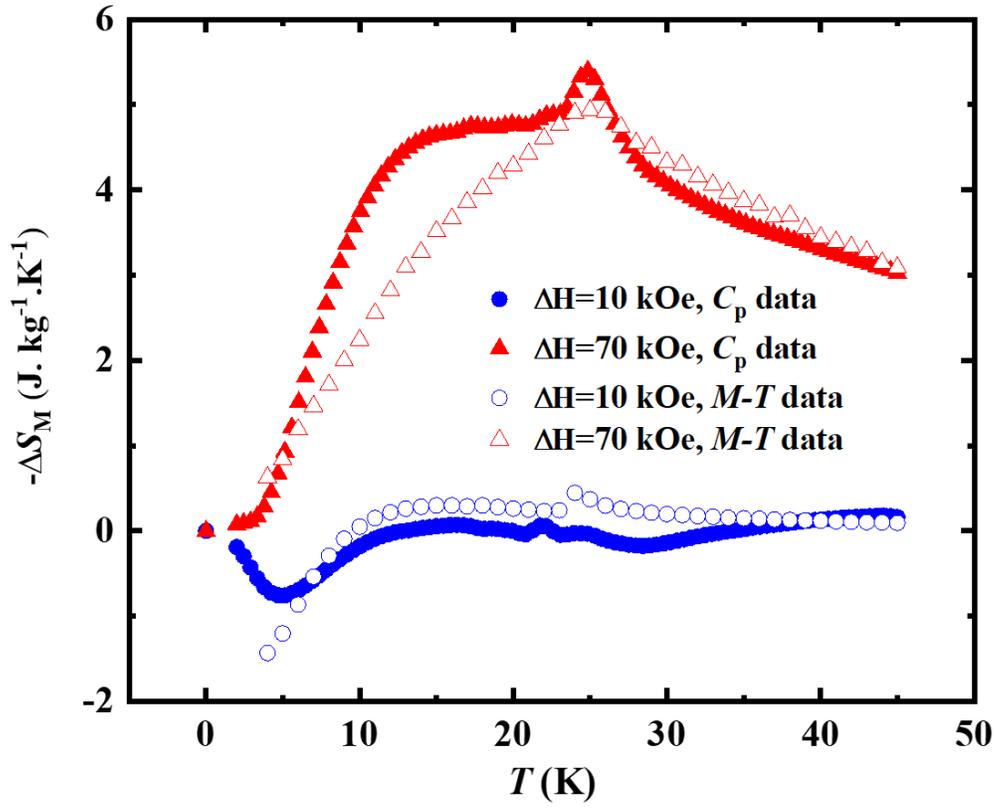

**Fig**. S1: Magnetic entropy of HoCrWO$_6$. The change in magnetic entropy, ΔSM, as function of temperature, *T* as calculated from specific heat (closed symbols) and magnetization data (open symbols).